\documentclass[aps,prb,twocolumn,floatfix]{revtex4}
\usepackage{epsf}
\usepackage{graphics}

\newcommand{\myscalebox}[1]{\scalebox{0.35}[0.35]{#1}}
\newcommand{\myscaleboxb}[1]{\scalebox{0.52}[0.52]{#1}}

\begin{document}


\title{Critical exponents of four-dimensional random-field Ising systems}
\author{Alexander K. Hartmann}
\email{hartmann@lps.ens.fr}
\affiliation{Department of Physics, University of California,
Santa Cruz CA 95064, USA}
\affiliation{Ecole Normale Sup\'erieure\\
24, Rue Lhomond,
75231 Paris Cedex 05, France}
\date{\today}

\begin{abstract}
The ferromagnet-to-paramagnet transition of the four-dimensional 
random-field Ising
model with Gaussian distribution of the random fields
is studied. Exact ground states of systems with sizes up to $32^4$ 
are obtained using graph theoretical algorithms. The
magnetization,
the disconnected susceptibility, the susceptibility and
 a specific heat-like quantity
 are calculated. Using finite-size scaling
techniques, the corresponding critical exponents are obtained:
$\beta=0.15(6)$,
$\overline{\gamma}=3.12(10)$, $\gamma=1.57(10)$ and $\alpha=0$ (logarithmic
divergence). Furthermore, values for
the critical randomness $h_c=4.18(1)$ and the correlation-length
exponent $\nu=0.78(10)$ were found. These independently obtained
exponents are compatible with all (hyper-) scaling relations and
support the two-exponent scenario ($\overline{\gamma}=2\gamma$).

\end{abstract}

\maketitle

\section{Introduction}

Phase transitions of pure systems
\cite{STAT-amit1984,STAT-cardy1996}
are already relatively well understood. The
critical behavior of all physical quantities can be described via
critical exponents. These exponents are related through 
(hyper-) scaling relations to each other, so that only {\em two} independent
exponents remain. In contrast, phase transitions  in systems with (quenched)
disorder \cite{young1998} exhibit many puzzles and are still far
from being understood. 

In theoretical physics, the random-field Ising
magnet (RFIM)  is a widely studied prototypical disordered system. 
It is  believed \cite{fishman1979,cardy1984}  to be in the same
universality class as the diluted antiferromagnet in a field, 
which can be studied
experimentally \cite{RFIM-belanger1998}. 

For a while it was thought \cite{aharony1976,young1977,parisi1979} 
that the critical behavior of the
$d$-dimensional RFIM is equal to that of the $d-2$ pure
ferromagnet. This would imply that the  $d=3$  RFIM exhibits
no ordered phase. This is not true, as has been shown later rigorously
\cite{bricmont1987}. In the meantime, a scaling theory
\cite{villain1985,bray1985,fisher1986} for the RFIM was developed,
where the dimension $d$ has been replaced by $d-\theta$ in the
hyper-scaling relations,
 $\theta$ (sometimes called also $y$) being a {\em third} independent
exponent, in contrast to the pure case. 
An alternative approach \cite{schwartz1985,schwartz1985b}
leads to the consequence that $\theta$ is not independent but related
to the exponent $\eta$ describing the divergence of the
(disconnected) susceptibility via $\theta=2-\eta$. Further evidence
for the existence of only two independent exponent was recently found
by high-temperature expansions\cite{gofman1993}. This was confirmed in
three dimensions by Monte Carlo simulations \cite{rieger1995b} and by
exact ground-state calculations \cite{alex-rfim3,rfim-c}.
But the exponents found in these works do not fulfill the
scaling relation $\alpha+2\beta+\gamma=2$, being
$\alpha,\beta,\gamma=\nu(2-\eta)$ the critical exponents for the specific heat,
the magnetization and the susceptibility, respectively. 
On the other hand, the results obtained in a different way
 in the most thorough ground-state study in $d=3$ 
so far \cite{mf01} indeed do not violate this scaling relation. 
Also, a modified scaling relation $\alpha+2\beta+\gamma=1$ was proposed 
\cite{nowak1998}, but the exponents obtained in
Refs. \onlinecite{rieger1995b,alex-rfim3,rfim-c} do not match the new relation
either. Hence, to obtain more insight into the
scaling behavior of the RFIM and to improve the knowledge of the
critical behavior of random systems, 
here the four-dimensional model is studied.

The RFIM Hamiltonian is given by

\begin{equation}
{\cal H}=-J\sum_{\langle i,j \rangle} S_i S_j - \sum_i (h_i+H) S_i ,
\label{eq:ham}
\end{equation}
where the $S_i = \pm 1$ are Ising spins, $J$ is the interaction energy
between nearest neighbors, $h_i\equiv h\epsilon_i$ 
is the random field and $H$ an
uniform external field. Here the case $H=0$ is studied, but small values
of the external field are used to determine the susceptibility.
The values $\epsilon_i$ are independently distributed according a
Gaussian distribution with mean 0 and standard deviation $1$,
i.e. the probability distribution is

\begin{equation}
P(h_i)=\frac{1}{\sqrt{2\pi}\, h}\exp\left(-{h_i^2 \over 2} \right) .
\end{equation}
Hence, $h_i=h\epsilon_i$ is Gaussian distributed with standard
deviation $h$.
We shall consider finite-dimensional lattices with periodic boundary
condition and $N=L^d$ spins. The results presented below are for $d=4$.

\begin{figure}
\begin{center}
\epsfxsize=0.9\columnwidth
\epsfbox{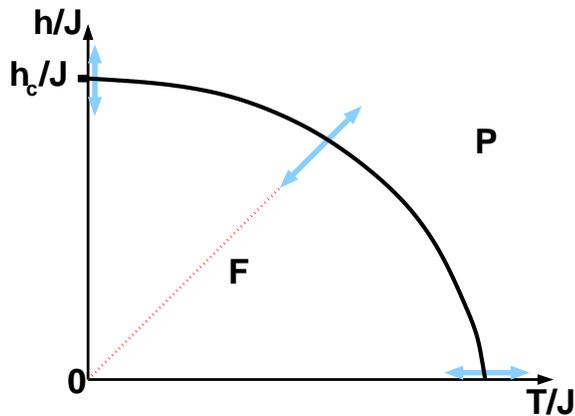}
\end{center}
\caption{A sketch of the phase boundary of the random field Ising model. The
ferromagnetic phase is denoted by ``F'' and the paramagnetic phase by ``P''.
The critical value of the random field at $T=0$ is denoted by $h_c$. The lines
with arrows at both ends indicate the path followed by varying $J$ for some
fixed value of $h$ and $T$.
}
\label{figPhase}
\end{figure}

The canonical phase diagram in zero external field ($H=0$) 
of the RFIM in higher than two dimensions 
is shown in Fig. \ref{figPhase}.
For low temperatures and small randomness, the ferromagnetic couplings
dominate, hence the system exhibits a long-range order. For higher
temperatures, where the entropy dominates, or for higher random fields,
where the spins are predominately aligned parallel to its local random
fields, the system is paramagnetic.

For random-field distributions which exhibit a maximum at $h=0$, such
as present in the Gaussian case, in mean-field theory
\cite{aharony1978} the phase transition
is second order along the whole phase boundary. Furthermore, the
renormalization group  flow along the phase boundary approaches the
$h=h_c$, $T=0$ fixed point \cite{fishman1979}, 
i.e. it controls the critical behavior. 
Therefore it can be expected  that the critical
behavior along the whole phase boundary is equal to that at
$h=h_c,T=0$, and all critical exponents can be obtained from the zero
temperature behavior. Working at
$T=0$ has the advantage that exact ground states of the RFIM can be
calculated using modern graph theoretical algorithms in polynomial
time\cite{angles-d-auriac1997b,rieger1998,alava2001,opt-phys2001}. 
This avoids equilibration problems often encountered in
Monte Carlo simulations and, even better, allows to study much larger
system sizes than before. 

The $T=0$ behavior of the four-dimensional RFIM has
been studied \cite{swift1997} using exact ground states up to size
$L=10$. 
In addition to the random  critical random field $h_c$, the exponents
 $\nu$ describing the divergence of the correlation length $\xi$ and
 $\beta$ for the magnetization were obtained. In this work, not only
much larger systems sizes up to $N=32^4$ are considered, but also 
 the critical exponents $\alpha$ for the specific heat
\cite{rfim-c},
$\gamma$ for the susceptibility and ${\overline{\gamma}}$
 for the
disconnected susceptibility are obtained {\em independently},
i.e. without using scaling relations. This in turn allows to test (hyper-)
scaling relations and to investigate the assumptions of two or three
independent critical exponents. The main results of this paper are
that the specific heat diverges logarithmically or maybe faster and
that the two independent exponent scenario is supported.

The rest of the paper is organized as follows. Next, the algorithms
used to calculated exact ground states are briefly explained. In the
main section all observables and results are presented. In the final
section the scaling relations are checked, the results discussed and a
summary given.

\section{Algorithms}

\label{sec:numerics}
To calculate the exact ground states at given randomness $h$ and field
$H$, algorithms
\cite{angles-d-auriac1997b,rieger1998,alava2001,opt-phys2001}
from graph theory
\cite{swamy,claibo,knoedel} were applied. To implement them,
some algorithms from the LEDA library \cite{leda1999} were utilized.

Here the methods are just outlined. More details can be found in the 
literature cited below or in the pedagogical presentation in 
Ref. \onlinecite{opt-phys2001}.
For each realization of the disorder, 
given by the external field $H$ and the values
$\{h_i\}$ of the random fields,
the calculation works by transforming the
system into a network \cite{picard1}, calculating the maximum flow
in polynomial time 
\cite{traeff,tarjan,goldberg1988,cherkassky1997,goldberg1998} and
finally obtaining the spin configuration $\{S_i\}$ from the values of
the maximum flow in the network. 
The running time of the latest maximum-flow methods has a peak
near the phase transition and
diverges\cite{middleton2002b,middleton2002}
 there like $O(L^{d+1})$
The first results of applying these algorithms to random-field
systems can be found in Ref. \onlinecite{ogielski}. In Ref.
\onlinecite{alex-rfim3} these methods were applied to obtain the exponents 
for the magnetization, the disconnected susceptibility and the
correlation length from ground-state calculations up to size $L=80$.
The most thorough study of the ground states of the 3d RFIM so far
is presented in Ref. \onlinecite{mf01}.
Other exact ground-state calculation of the 3d model can be found
in Refs. \onlinecite{angles-d-auriac1997,sourlas1999,nowak1998}.
These techniques have also already applied to small four-dimensional
systems \cite{swift1997}.

Since the algorithms work only with integer values for all parameters,
a value of $J=10000$ was chosen here,  
and all values were rounded to its nearest integer
value. This discreteness is sufficient, as shown in Ref. \onlinecite{mf01}. 
All results are quoted relative to $J$ (or assuming $J\equiv 1$).

Note that in cases where the ground-state
is degenerate \cite{foot-degenerate}
it is possible to calculate all the ground-states in one
sweep \cite{picard2}, see also
Refs. \onlinecite{alex-daff2,bastea1998}. For the
RFIM with a Gaussian distribution of fields, the ground state is
non-degenerate, except for a 
two-fold degeneracy at certain values of the randomness, where the
ground state changes, 
so it is sufficient to calculate just one ground state.

\section{Results}

In this work,  exact ground states for system sizes $L=4$ to
$L=32$ for different values of the randomness $h$ and with 4 different
values $H=0,H_L,2H_L,4H_L$ (only $H=0$ for
$L=32$) were calculated. Near the transitions, an average over the
disorder with the number $N_{\rm samp}$ of samples ranging from
 3200 ($L=32$) up to 40000 ($L=4$) was performed, less
samples were used away from the critical point, since the fluctuations
are small outside the critical region. For details, see
Tab. \ref{tab:hvals}. 

\begin{table}[ht]
\begin{tabular}{l|rr}
L & $N_{\rm samp}$ & $H_L$ \\\hline
4 & $40000$ & $0.025$ \\
6 & $20000$ & $0.02$  \\
8 &  $7100$ &  $0.012$ \\
12 & $8500$ & $0.006$ \\
16 &  $4000$ &  $0.003$ \\
24 &  $10000$ & $0.0015$ \\
32 &  $3200$ &  $-$ 
\end{tabular}
\caption{The maximum number of samples $N_{\rm samp}$ used, and sizes of
smallest non-zero uniform field $H_L$, for each system size $L$. As discussed
in the text, the number of samples used was larger in the vicinity of the peaks
in the susceptibility and specific heat than elsewhere. 
}
\label{tab:hvals}
\end{table}

We first concentrate on the case $H=0$. The simulations with $H>0$
were done to obtain the susceptibility, see below.

\begin{figure}[htb]
\begin{center}
\myscalebox{\includegraphics{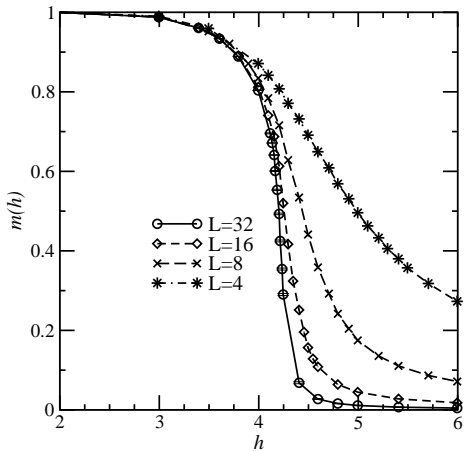}}
\end{center}
\caption{Average magnetization as a function of random-field strength
$h$. For clarity of the plot, only $L=4,8,16,32$ are shown. Error bars
(shown for $L=32$) are much smaller than symbol
sizes. Lines are
guides to the eyes only.
}
\label{figMag}
\end{figure}

As, already mentioned, when studying {\em one} 
single finite sample of the disorder
$\{h\epsilon_i\}$ as a function of $h$, the ground state changes only at
certain discrete values
$h^{(1)},h^{(2)},\ldots,h^{M(\{\epsilon_i\})}$. Hence, quantities like the
magnetization are stepwise constant as a function of $h$. This
discreteness vanishes, when averaging over the disorder.

In Fig. \ref{figMag} the average magnetization per spin
\begin{equation}
m \equiv \left[ |M \right|]_h \equiv 
\left[\left|\frac{1}{N}\sum_i S_i \right|\right]_h 
\end{equation}
is shown as a function of the randomness $h$ for system sizes
$L=4,8,16,32$.  The average over the disorder is denoted by
$[\ldots]_h$,  which is carried
out (approximately) by repeating the calculation for $N_{\rm samp}$
independent realizations (samples) of
the random fields $\{h\epsilon_i\}$. 
As expected, for small randomness the ground state is
ferromagnetically ordered and disordered for large values of $h$. The
curves become steeper with increasing $h$, indicating a phase
transition near $h=4.2$.

\begin{figure}[htb]
\begin{center}
\myscalebox{\includegraphics{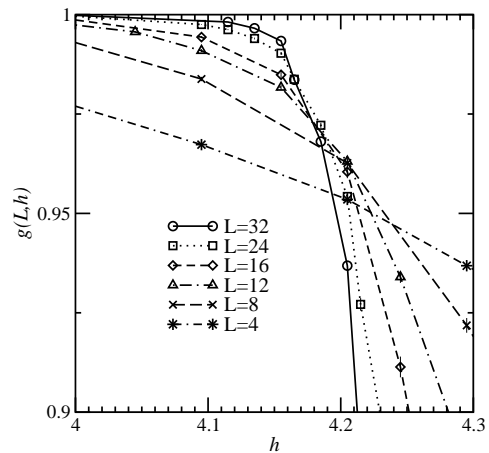}}
\end{center}
\caption{Binder parameter $g(L,h)$ as a function of the randomness $h$
for different system sizes $L$. For clarity of the plot, only $L=6$
is omitted.  Error bars are smaller than symbol
sizes. Lines are
guides to the eyes only.
}
\label{figBinder}
\end{figure}

To study the transition more detailed, the Binder parameter
\cite{binder81,bhatt85}
\begin{equation}
g(L,h)\equiv\frac{1}{2}
\left( 3-\frac{[\langle M^4\rangle]_h }{[\langle M^2\rangle]_h^2}\right)
\end{equation}
is calculated, $\langle \ldots \rangle$ being the thermal disorder,
(which is trivial at $T=0$ if the ground state is non degenerate). 
The idea behind the definition of this quantity is 
that in the thermodynamic limit, the distribution of the order parameter
should converge towards a delta function (with $g(L,h)=1$) 
in the ordered phase  and to a Gaussian distribution $(g(L,h=0))$ in the
disordered phase. 
The scaling theory for 2nd order phase transitions assumes, 
that the finite-size behavior is governed by the
ratio $L/\xi$, $\xi$ being the correlation length. At the
critical point, where the correlation length is infinite, the
parameters for different system sizes should intersect, since $L/\xi$
is sero for all sizes $L$.
In Fig. \ref{figBinder} the result for the 4d RFIM is shown, please
note the enlarged scale.

Indeed all curves intersect near $h=4.2$, compatible with the result
for the magnetization.  A systematic shift can be observed, which
decreases with growing system size, and is due to finite-size
effects in small systems. From the intersections of sizes $L\ge 8$, a
value for the critical randomness of $h_c^{\rm binder}=4.18(2)$ is
estimated. Due to the finite-size effect observed here, 
small system sizes will be
excluded from subsequents fits if they don't match the leading behavior.

To observe the specific heat singularity, here the bond energy $E_J$ is
studied\cite{rfim-c}, 
given by
\begin{equation}
E_J \equiv {\partial F \over \partial J}
= -{1 \over N} \sum_{\langle i,j\rangle} \langle S_i S_j \rangle,
\label{eq:EJ}
\end{equation}
where the sum is over nearest-neighbor pairs. $E_J$ has an energy-like
singularity in the vicinity of the phase boundary. 
Having differentiated {\em analytically}
with respect to $J$, now $J=1$ is set, and $T=0$ considered 
only.  A specific heat-like quantity is obtained by
differentiating $E_J$ {\em numerically}\/ 
with respect to the random field $h$. A first-order
finite difference is used to determine the derivative
of $E_J$ numerically and, since this is a more accurate representation of the
derivative at the midpoint of the interval
than at either endpoint, the ``specific heat'',
$C$, at $T=0$ is defined to be
\begin{equation}
C\left({h_1 + h_2 \over 2}\right) \equiv {[E_J(h_1)]_h - [E_J(h_2)]_h \over
h_1 - h_2} ,
\label{eq:sh}
\end{equation}
where $h_1$ and $h_2$ are two ``close-by'' values of $h$.
A sufficiently fine mesh of random-field values is chosen such 
that the resulting
data for $C$ is smooth. Error bars are obtained by determining the ``specific
heat'' from the corresponding finite difference as in Eq.~(\ref{eq:sh})
for each sample separately, and
computing the standard deviation. The error bar is, as usual, the standard
deviation divided by $\sqrt{N_{\rm samp}-1}$.

\begin{figure}[htb]
\begin{center}
\myscalebox{\includegraphics{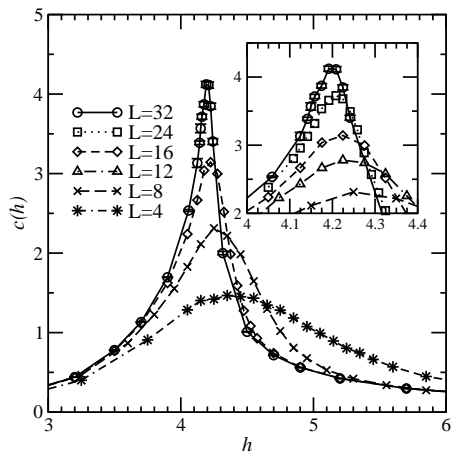}}
\end{center}
\caption{Specific heat-like quantity $C=\partial^2 F/\partial h
\partial J$ as a function of the randomness $h$ for $L=4,8,16,32$.
Error bars are only shown for $L=32$. For the other system sizes the
error bars are even smaller. 
The inset shows the region near the peaks enlarged for $L\ge 8$. Lines
are guides to the eyes only.
}
\label{figHeat}
\end{figure}

In Fig. \ref{figHeat} the numerical results are exposed. The
``specific heat'' exhibits clear peaks, which grow in system size and
move slightly left. The number of samples used
is larger near the peak to compensate for the greater sample to sample
fluctuations in this region. 

In a finite system, finite-size scaling predicts for the singular part
$C_{\rm s}$ of the specific heat
\begin{equation}
C_{\rm s} \sim L^{\alpha/\nu}
\widetilde{C}\left( (h - h_c) L^{1/\nu}\right)  ,
\label{eq:scale:heat}
\end{equation}
where $\nu$ is the correlation length exponent. The
``specific heat'' peak will occur when the
argument of the scaling function $\widetilde{C}$ takes some value, $a_1$ say,
so 
the peak position $h^{*}(L)$ varies as
\begin{equation}
h^{*}(L)-h_c \approx a_1 L^{-1/\nu} ,
\label{eq:shift:max}
\end{equation}
and the value of the singular part of the ``specific heat'' at
the peak 
varies as
\begin{equation}
C_{\rm s}^{\rm max}(L) \sim  L^{\alpha/\nu} .
\end{equation}

\begin{figure}[htb]
\begin{center}
\myscalebox{\includegraphics{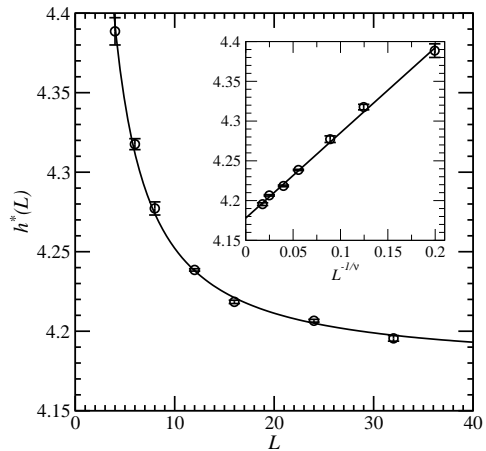}}
\end{center}
\caption{A plot of the random field where the ``specific
heat'' attains its maximum, as a function of system size $L$. The
solid line shows a fit to the function 
$h^{*}(L)=h_c + a_1 L^{-1/\nu}$ with $h_c=4.18$,
$1/\nu=0.78$, and $a_1 = 1.34$.
The inset shows the data as a function of $L^{-1/\nu}$.
}
\label{figDeltaCL}
\end{figure}

For each system size,   parabolic fits were performed in
 the region of the peak
to obtain $h^{*}(L)$ and the height of the peak, $C^{\rm max}(L)$.
The shift of the maximum according to Eq. (\ref{eq:shift:max}) can be used
to estimate the infinite-size critical strength 
of the random field, $h_c$ and the
correlation-length exponent $\nu$.
The best fit gives
\begin{equation}
h_c = 4.182 \pm 0.006, \qquad 1/\nu=1.28 \pm 0.14 ,
\label{eq:hc-sh}
\end{equation}
see Fig. \ref{figDeltaCL}.
The probability $Q$ was determined that the value of 
$\chi^2=\sum_{i=1}^N (\frac{y_i-f(x_i)}{\sigma_i})^2$, 
with $N$ data points ($x_i,y_i\pm\sigma_i$) fitted
to the function $f$, is worse than in the current 
fit \cite{numrec1995} to quantify the quality of the fit. Here a value
of $Q=0.06$ was obtained , which is not very good. 
The reason is, that the error bars
used for the positions and heights of the maxima were only the
statistical error bars obtained from the fit of the parabolas, and
which are often surprisingly small. Systematic errors, resulting from
the fact that the peaks are in fact not parabolic, are not included in
this way. Therefore, fits to a fourth order polynomials were tried,
but the results turned out to be very unreliable and the fits very
unstable against the change of the window over which the data was
fitted. Hence, the parabolic fits were kept, were these effects were
smaller, and the final value quoted is $h_c=4.18(1)$.

Next, the singular behavior of $C$ is determined by analyzing, how
the peak value $C^{\max}$ scales with $L$.
Please note \cite{rfim-c} that at $T=0$ the singular behavior $C$ is equal
to the singular behavior of $C^{\prime}=-
{\partial^2 F}/{\partial h^2}$, since $C=-C^{\prime}h/J$ at $T=0$.

\begin{figure}[htb]
\begin{center}
\myscalebox{\includegraphics{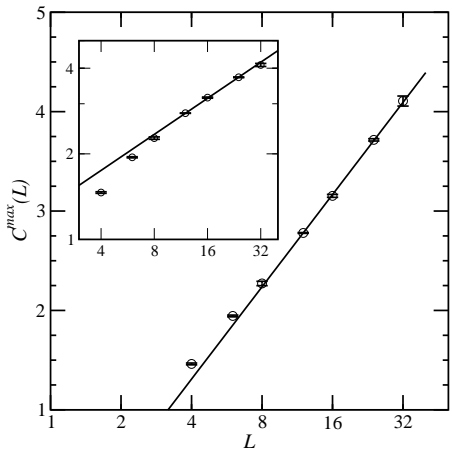}}
\end{center}
\caption{The maximum $C^{\rm max}$ of the ``specific heat'' as a
function of system size with logarithmically scaled $L$-axis. The
lines shows the function $a+b\log L$ with $a=-0.53$ and $b=1.33$.
The inset shows the data in a double-logarithmic plot, the solid line
being the function $a_2L^{\alpha/\nu}$ with $\alpha/\nu=0.419$ and
$a_2=0.98$. }
\label{figHeatMaxL}
\end{figure}

The first hypothesis is that the ``specific heat'' diverges
logarithmically, as found in experiments for three-dimensional diluted
antiferromagnets in a field \cite{belanger}. Hence, the values of
$C^{\rm max}$ are fitted ($L\ge 8)$ to a function of the form
\begin{equation}
C^{\rm max}(L) = a + b \log L ,
\label{eq:logfit}
\end{equation}
where the constant term $a$ comes partly from the regular piece of the
``specific heat''. The quality $Q=0.23$ of the fit is fair. In
Fig. \ref{figHeatMaxL} the ``specific heat'' is shown in a logarithmic
plot together with the fit function (\ref{eq:logfit}). For larger
system sizes the datapoints follow very nicely a straight line,
suggesting that $C$ indeed may diverge logarithmically. Please note
that the ``specific heat'' peaks seem to be rather symmetrical, which means
that the amplitudes $A_{+},\,A_{-}$ (for $h>h_c\,h<h_c$) of $C$ are
almost equal. This is exactly what one expects in the case of a
logarithmic divergence \cite{privman1991}.

Also the possibility of a algebraic divergence was tested. For this
purpose the data was fitted ($L\ge 8$) to the function
\begin{equation}
C^{\rm max}(L) = a_2L^{\alpha/\nu} ,
\label{eq:shmax}
\end{equation}
resulting in $\alpha/\nu=0.419(9)$ and $a_2=0.98(2)$, the result is
shown together with the data in a double logarithmic plot in the inset
of Fig. \ref{figHeatMaxL}. The quality $Q=9\times 10^{-3}$ of the fit 
is very bad. To check whether this is an effect of including too small
system sizes into the fit, also a fit using only system sizes $L\ge 12$ was
performed, resulting in $\alpha/\nu=0.416(6)$, $a_2=0.99(1)$ and
$Q=0.16$ which is much better. Since for $L\ge 12$ the logarithmic fit
has also a better quality $Q=0.55$, and because the negative curvature
in the double-logarithmic plot is more pronounced than a possible
positive curvature in the single-logarithmic plot, 
still a logarithmic divergence seems more likely from this data, i.e.
\begin{equation}
\alpha=0   ,
\end{equation}
but an algebraic behavior with a small exponent cannot be excluded.

\begin{figure}[htb]
\begin{center}
\myscalebox{\includegraphics{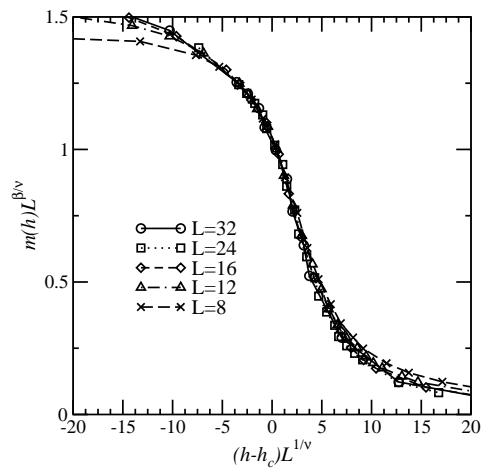}}
\end{center}
\caption{Scaling plot of the rescaled absolute value
$m(h)L^{\beta/\nu}$ magnetization as a function of of
$(h-h_c)L^{1/\nu}$ with $h_c=4.18$,  $1/\nu=1.28$ and $\beta/\nu=0.17$. 
 Error bars are smaller than symbol
sizes. Lines are guides to the eyes only. } 
\label{figMagScale}
\end{figure}

Next, the critical behavior of the magnetization is studied. The
predictions from finite-size scaling is that near the critical point
\begin{equation}
m(h)=L^{-\beta/\nu}\tilde{m}((h-h_c)L^{1/\nu})  .
\end{equation}
This means, that by plotting $m(h)L^{\beta/\nu}$ 
against $L^{1/\nu}(h-h_c)$ with correct
parameters $h_c,\nu$ and $\beta/\nu$, 
the data points for different system sizes should
collapse onto a single curve near $(h-h_c)=0$. The values $h_c=4.18$
and $1/\nu=1.28$ from above were used. With 
\begin{equation}
\beta/\nu=0.17(5)
\end{equation}
the best collapse of the data for $L\ge 8$ was obtained, which is
presented in Fig. \ref{figMagScale}.

\begin{figure}[htb]
\begin{center}
\myscaleboxb{\includegraphics{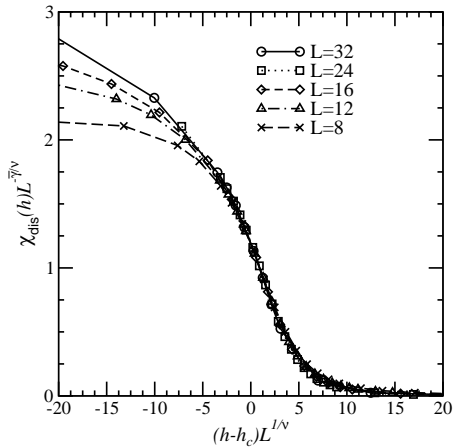}}
\end{center}
\caption{Scaling plot of the rescaled disconnected susceptibility
$\chi_{\rm dis}(h)L^{\overline{\gamma}/\nu}$  as a function of of
$(h-h_c)L^{1/\nu}$ with $h_c=4.18$,  $1/\nu=1.28$ and 
$\overline{\gamma}/\nu=3.63$. Error bars are smaller than symbol
sizes. Lines are guides to the eyes only. 
}
\label{figDisScale}
\end{figure}

The singular behavior of the disconnected susceptibility 
\begin{equation}
\chi_{\rm dis} \equiv L^d\left[ M^2 \right]_h
\end{equation}
can be obtained in an analogous way to the magnetization. The
following scaling behavior is assumed:
\begin{equation}
\chi_{\rm dis}(h)=L^{\overline{\gamma}/\nu}
\tilde{\chi}_{\rm dis}((h-h_c)L^{1/\nu})  .
\end{equation}
From collapsing the date curves for $L\ge 8$, using $h_c=4.18$
and $1/\nu=1.28$, a value of 
\begin{equation}
\overline{\gamma}/\nu=3.63(0.05)
\end{equation}
 was found. The scaling plot is shown in
 Fig. \ref{figDisScale}. Please note that the scaling behavior of
 $h<h_c$ is worse than for $h>h_c$ (also, to a lesser extent, in
 Fig. \ref{figMagScale}). The reason is that for smaller fields the
 systems quickly become fully ordered ($m=1$), i.e. scaling does not hold.

Finally, the susceptibility and its related critical exponent $\gamma$
is determined. This is done by considering the response to a small
uniform external field $H>0$.  For each realization, the sign of $H$
is chosen in the direction of the magnetization of the ground
state. This prevents the 
whole system from flipping when applying a magnetic field to a system
which is almost ferromagnetically ordered, which would mimic a false
high susceptibility. The scaling behavior of the
magnetization should not be affected by this choice.

The ground state for each realization and each value of $h$ is
calculated for $H_n=0,H_L,2H_L,4H_L$. Near $H=0$, the data points can be fitted
very well with a parabola, the coefficient of the linear term gives
the zero field susceptibility 
\begin{equation}
\chi=dm/dH|_{H=0} .
\end{equation}
 To cope for the
expected strong increase of $\chi$ with the system size, $H_L$ must
strongly decrease with $L$, in the order of the expected increase.
For details see
Ref. \onlinecite{rfim-c}, the values of $H_L$ are shown in
Tab. \ref{tab:hvals}. Then, for each system size,
a fit to a parabola through the four data points for the {\em average}
magnetization $m(H_n)$ is performed. To estimate
the error, a jackknife analysis \cite{jackknife} was used,
in which the results for the magnetizations
(for each system size and each strength of
the disorder) is divided into $K$ blocks, the average values
calculated $K$ times,
each time omitting one of the blocks, and then $K$ fits are performed. The
error bar is estimated from the variance of the $K$ results for 
the linear fitting parameter. Here $K=50$ was used and checked that
the result does not depend much on the choice of $K$.

\begin{figure}[tb]
\begin{center}
\myscalebox{\includegraphics{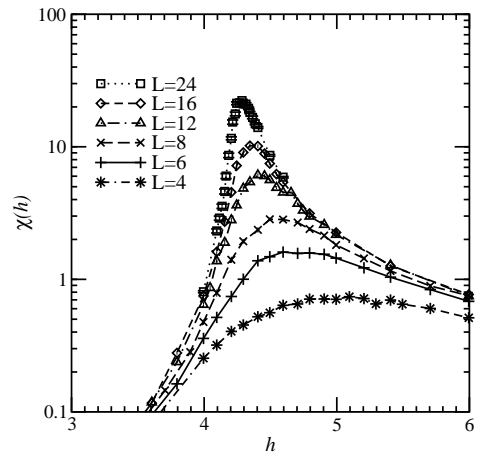}}
\end{center}
\caption{Susceptibility $\chi$ as a function 
of the random-field strength $h$ 
for system sizes $L=4,6,8,12,16,$ and $24$. Error bars (shown for
$L=24$) are smaller than symbol sizes. Lines are guide to the eyes only. 
}
\label{figSusz}
\end{figure}

\begin{figure}[htb]
\begin{center}
\myscalebox{\includegraphics{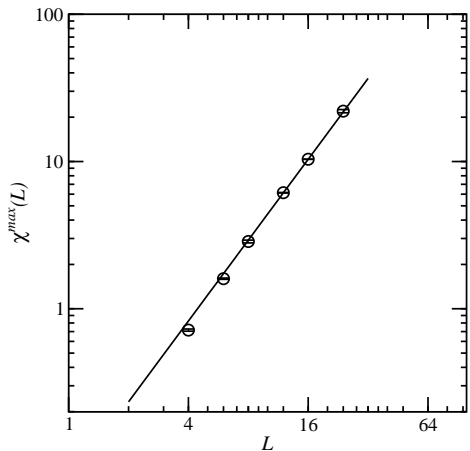}}
\end{center}
\caption{The maximum $\chi^{\rm max }$ of
the susceptibility as a function of system 
size $L$ in a double logarithmic plot. 
The solid line represents a fit to the
function $\chi^{\rm max}(L)=a_3 L^{\gamma/\nu}$, for sizes
$L \ge 8$ yielding $\gamma/\nu = 1.82$ and $a_3=0.066$. }
\label{figSuszHeightL}
\end{figure}

The susceptibility $\chi$ as a function of $h$ is presented in
Fig. \ref{figSusz} for selected system sizes. 
It is seen that the height of the peak grows much faster than for the
``specific heat''. For the susceptibility, the following scaling behavior
is expected:
\begin{equation}
\chi(h)=L^{\gamma/\nu}\tilde{\gamma}((h-h_c)L^{1/\nu})  .
\end{equation}
To analyze the divergence of $\chi$, again parabolas were fitted 
to the data points near the peak to obtain the positions
$h^{*}(L)$ and $\chi^{\rm max}(L)$
of the maximum.
By fitting the data for $L \ge 8$ to
a function $\chi^{\rm max}(L)=a_3 L^{\gamma/\nu}$, where $\gamma$ describes the
decay of the ``connected'' correlations at criticality, the following
values were obtained ($Q=0.54$)
\begin{equation}
\gamma/\nu = 1.82 \pm 0.01 , 
\label{eq:eta}
\end{equation}
see Fig. \ref{figSuszHeightL}.

It was also tried to fit the positions $H^{*}(L)$ of the maxima of the
susceptibility to a scaling function of the from
(\ref{eq:shift:max}). But the quality of the fit was very bad
($Q<10^{-7}$)  for all
ranges $[L_{\rm min},24]$ considered and the result for $h_c$ was
always too large ($h_c>4.26\pm 0.03)$. This value is not only not
compatible with the result from the ``specific heat'' $C$, but in clear
contradiction to the Binder parameter (see Fig. \ref{figBinder}).
The reason is probably the
stronger finite-size dependence of the position of the susceptibility
peak in comparison to the position of the ``specific heat'' peak,
please compare Figs. \ref{figHeat} and \ref{figSusz}. Also,
due to the four times higher numerical effort to determine $\chi$,
only simulations for $L\le 24$ were performed. Furthermore, the peaks
for the susceptibility are much broader than for the ``specific
heat'', hence it is harder to determine the position of the peak
precisely. Thus, the result $h_c=4.18(1)$ from the previous measurement
was kept.

\section{Discussion}
Using graph-theoretical algorithms, exact ground states of 
random-field Ising systems were calculated in polynomial time. Using the LEDA
library, system sizes up to $N=32^4$ could be considered.

All critical exponents, describing the order-to-disorder transition
at $h_c$ were calculated independently. The following values, using
$\gamma=\nu(2-\eta)$ and $\overline{\gamma}=\nu(4-\overline{\eta})$, for the
values of the critical field $h_c$ and the exponents were obtained:
\begin{equation}
\begin{array}{l@{\:=\:}r@{}ll@{\:=\:}l}
h_c &  4.18  &  \pm 0.01, &  {\nu} & 0.78 \pm 0.10 \\
\alpha & 0  &  &  \beta & 0.13 \pm 0.05 \\
\eta & 0.18 & \pm 0.01 &  \overline{\eta} & 0.37 \pm 0.05 \\
\gamma & 1.42 & \pm 0.20 &  \overline{\gamma} & 2.83 \pm 0.50 .
\end{array}
\end{equation}
Please note that the values for $\beta$, $\gamma$ and $\overline{\gamma}$ carry a
large error bar due to the uncertainty in the correlation-length
exponent $\nu$.

The results for the critical field, the correlation exponent and the
exponent of the magnetization are compatible with values found formerly
via the exact ground state calculations of small
systems\cite{swift1997}  up to $L=10$,
where $h_c=4.17(5)$, $\beta=0.13(2)$ and $\nu=0.8(1)$
 were obtained. The result for the susceptibility exponent is
compatible with the result $\gamma=1.45(5)$ which was 
found in a high-temperature
expansion \cite{gofman1993}. The results obtained here are also
compatible with the exponents obtained recently \cite{middleton2002}
by the use of exact ground states as well but by evaluating mainly other
quantities like distributions of domain-wall energies or fractal
properties of domain walls. The main difference is that the results
from Ref. \onlinecite{middleton2002} have a slight preference for the
exponent $\alpha$ of the ``specific heat'' to be positive
but small.

Next, the validity of the scaling relation for the ``specific
heat'' is checked. For the Rushbrooke equality 
\begin{equation}
\alpha+2\beta+\gamma = 2
\end{equation}
one gets  $\alpha+2\beta+\gamma=1.68(30)$, which is not very good but still
within the error bars almost fulfilled. In case the algebraic divergence is
taken ($\alpha=0.33(5)$) a value of $\alpha+2\beta+\gamma=2.11(35)$ is
obtained, which fulfills the equation better.

The deviation of the hyper-scaling relations from the pure case is
obtained by replacing $d$ by $d-\theta$, 
with the exponent $\theta=2-\overline{\eta}+\eta=1.81(6)$. E.g.
the hyper-scaling relation 
\begin{equation}
2-\alpha=\nu(d-\theta)
\end{equation}
is also fulfilled within error bars ($\nu(d-\theta)=1.70(30)$), again
the case were it is assumed that the ``specific heat'' diverges faster
than logarithmic matches the relation better.

Finally, we turn to the question whether there are two or three
independent exponents. In the case of two exponents, the
Schwartz-Soffer equation \cite{schwartz1985} 
\begin{equation}
\overline{\gamma}=2\gamma
\end{equation}
holds, which is compatible with the result found here. Hence, the
two-exponent scenario is supported.

To summarize, all critical exponents of the four-dimensional RFIM were
determined independently. All ``classical'' 
(hyper-) scaling relations are fulfilled
and the two-independent-exponents scenario is supported. The scaling
relation proposed in Ref. \onlinecite{nowak1998} seems not to be fulfilled.
The largest uncertainty in the results presented here is in the value
of $\alpha$, whether $\alpha=0$ or $\alpha>0$ and small. To finally
decide this question, much larger system sizes must be studied,
which are currently out of reach.

\begin{acknowledgments}

The author thanks A.P. Young for hosting him at the University of
California  Santa Cruz, where the work was performed, 
many helpful discussions and critically
reading the manuscript. The work has benefited much from discussions
with D. Belanger, R. K\"uhn, A.A. Middleton, M. Moore and N. Sourlas.
The simulations were performed on a Beowulf Cluster
at the Institut f\"ur Theoretische Physik of the Universit\"at
Magdeburg and   at the Paderborn Center
for Parallel Computing both in Germany.
Financial support was obtained from the DFG (Deutsche 
Forschungsgemeinschaft)
under grant Ha 3169/1-1.
\end{acknowledgments}

\end{document}